# Programmable calculus operations in electromagnetic space using space-time-coding metasurface

Hao Tian Shi[1,†], Lei Zhang[1,†,*], Rui Yuan Wu[2,*], Yi Ning Zheng[1], Xiao Qing Chen[1], Yuanzhe Li[1,3], Shi He[1], Jun Wei Wu[1], Qiang Cheng[1], and Tie Jun Cui[1,*]

[1] *State Key Laboratory of Millimeter Waves, Southeast University, Nanjing 210096, China*

[2] *College of Information Science and Engineering, Hohai University, Nanjing 211100, China*

[3] *National Key Laboratory of Automatic Target Recognition, College of Electronic Science and Technology, National University of Defense Technology, Changsha 410073, China*

[†] *These authors contributed equally to this work.*

[*] Corresponding authors: Email: lzseu@seu.edu.cn; ruiyuanwu@126.com; tjcui@seu.edu.cn.

**Abstract**

With the rapid advancement of metasurfaces and the increasing demand for programmable metasurfaces to simplify information systems, wave-based computation using metasurfaces has emerged as an attractive research topic. To facilitate the mathematical operations in electromagnetic (EM) space, here we propose a space-time coding metasurface (STCM) system capable of directly performing calculus operations on the spatial energy distributions of EM waves. By exploiting harmonic characteristics induced by time-varying coding, the responses of meta-atoms at specific harmonics can be flexibly controlled, which enables the metasurface system to address more complex tasks. Owing to its programmability, the STCM can dynamically switch functions in real time to accommodate different calculus tasks. To fully leverage the capability of STCM, we not only present the space-time coding sequences for differentiation and integration of EM waves, but also develop and numerically simulate the space-time coding sequences that can independently and simultaneously implement different calculus operations on the same incident EM waves. To experimentally validate the feasibility of the EM calculus operations, proof-of-concept experiments are conducted using a programmable 2-bit STCM. Good agreements among the theory, numerical simulations, and experiments confirm the feasibility of performing calculus operations in the EM space and demonstrate the broad application prospects of STCM in EM wave manipulations, wireless communications, and signal processing.

# Introduction

The proposal of programmable electromagnetic (EM) metasurfaces [1] provides a flexible and powerful approach for manipulating EM waves, which can be widely used in complex EM regulation [2-4], wireless communication [5-8], and sensing [9-11]. With the development of EM information theory [12,13], using metasurfaces to directly characterize and process information has become an appealing topic, and one of the key challenges is the direct implementation of mathematical operations on spatial EM waves using programmable metasurfaces. Compared with the digital computation that relies on complex computing hardware and a large amount of storage resources, performing mathematical operations directly on spatial EM waves is an analog computing paradigm with quasi-speed-of-light processing capability [14-16], which can directly process the EM signal in physical space and reduce the signal processing complexity in the digital hardware.

The convolution [17] and addition [18] operations on coding metasurfaces have provided both theoretical foundations and feasible solutions for directly computing with the EM waves. Although convolution and addition operations are widely applicable to functions such as beamforming and beam steering, the traditional programmable metasurfaces are constrained by a limited number of coding states, which makes it difficult to perform more complex mathematical operations. One possible solution to overcoming this limitation is the use of stacked programmable metasurfaces. Through careful structural design and coding patterns, the stacked metasurfaces can realize an all-EM-based artificial neural network through the diffraction [19], thereby enabling the approximation of various mathematical operations. To date, the stacked metasurfaces have been explored in diverse applications, including direction of arrival (DOA) estimation [20], computational image reconstruction [21], and intelligent sensing [22]. However, the programmable stacked metasurfaces suffer from inherent drawbacks, such as the increased system profile and complicated control architectures. Another possible solution to the limited coding states is the use of passive metasurfaces with fixed physical configurations. The phase and amplitude responses of a passive metasurface can be flexibly engineered through the geometric structure of meta-atoms, and recent studies have shown experimentally complex functionalities such as solving linear equations [23] and performing the calculus operations [24-25] on spatial EM waves. Nevertheless, once fabricated, the function of passive metasurface is fixed, which restricts it to a limited category of problems, limiting its capability for multifunctional and adaptive operations.

One promising solution to the aforementioned limitation is the adoption of space-time coding metasurfaces (STCMs) [26] for EM mathematical operations. By introducing temporal modulation into the coding scheme, the space-time coding metasurfaces can precisely control EM waves through an increased degree of time complexity. With carefully designed periodic coding sequences, the amplitude and phase responses at a specific harmonic frequency can be nearly continuously tuned. Owing to the space-time coding strategy and harmonic modulation characteristic, the space-time coding metasurfaces are well-suited for realizing a wide range of complex functionalities. Numerous studies have been explored on the space-time coding metasurfaces, such as coding optimization [27,28], EM manipulation [29-34], signal processing [33-37], and wireless communications [38-43], demonstrating their broad application potentials in scattering and radiation manipulation [44-49], hardware diagnostics [50,51], and imaging [52,53] in radio-frequency technology. Thus, employing the space-time coding metasurface provides an effective and versatile approach for implementing complex mathematical operations and reaching programmable functional features.

To verify the effectiveness of STCM in programmable EM mathematical operations, we design a wave-based programmable calculus operation platform based on a programmable 2-bit STCM with a specially designed space-time coding strategy. The 2-bit coding feature provides sufficient freedom for tailoring harmonic responses, while the programmability enables the multifunctional capability for implementing complex functions. For the proposed STCM, coding sequences corresponding to different functions, including differential and integral operations, are optimized and preloaded in the FPGA module, enabling the STCM to perform various functions according in response to different control instructions. Inspired by the convolution theory of metasurface and Fourier relationships between the near-field and far-field EM wave distribution, we carefully optimized several sets of space-time coding sequences, which can realize differential and integral operations in the EM space. Furthermore, by exploiting the harmonic resources, we design a space-time coding sequence that can perform the differential operation at the first harmonic and integral operation at the second harmonic simultaneously for the same incident wave. Good agreement between the design objectives and measured results demonstrates satisfied performance of the STCM platform in implementing EM mathematical operations, highlighting its broad application potentials in EM manipulation, signal processing, and imaging. In contrast to the previous study on the metasurface-based mathematical operations relying solely on one harmonic frequency [54] and only focused on theoretical analysis and numerical simulations, this work not only realizes a realistic STCM prototype for programmable EM calculus operations but also fully exploits the multi-harmonic resources to enhance functionality.

**Framework of metasurface-based calculus platform**

**Figure 1** illustrates the structural design of the proposed STCM-based calculus platform, along with a schematic representation of implementing differential operators. The STCM calculus platform is operated by exploiting two Fourier transform relationships, namely the Fourier relationships between the time-coding sequences and frequency-domain harmonics, and the Fourier relationships between the near-field responses and far-field scattering patterns. By leveraging these two Fourier-transform mechanisms, the calculus operations on incident EM waves can be effectively realized.

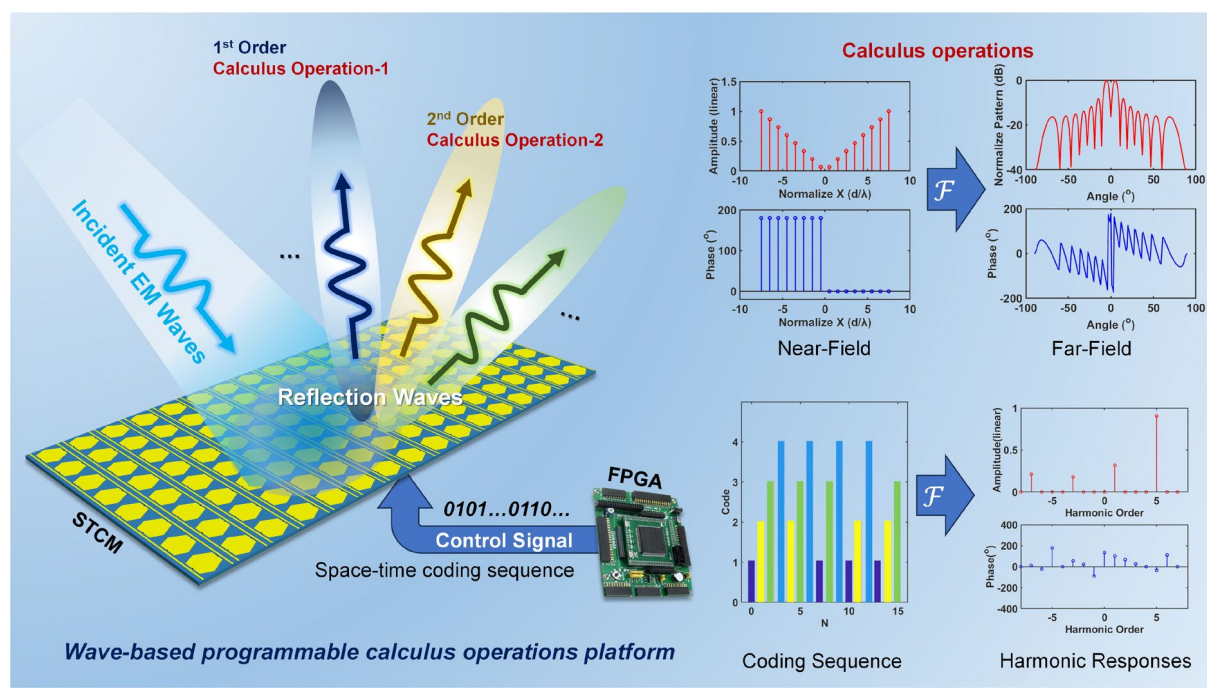

**Figure 1. The schematic illustration of the wave-based programmable calculus operations platform that can conduct different calculus operations in the +1st and +2nd harmonics using STCM.** STCM is controlled by an FPGA, and can characterize the calculus operators through different space-time-coding sequences and realize direct calculus operations towards the incident EM wave.

The Fourier transform relation between the time-coding sequence and harmonic response provides a promising solution for the arbitrary manipulations of the amplitude and phase responses at a specific harmonic frequency, thereby enabling the characterization of calculus operators on a metasurface. For a time-coding sequence of a given length $L$, the relationship between the coding sequence $c_n$ and the harmonic response $\Gamma_k$ can be expressed as [26]:

$$\Gamma_k = \sum_{n=1}^{L} c_n e^{-j\frac{k2\pi n}{L}} \tag{1}$$

The Fourier transform relation between the near-field EM response and the far-field pattern enables the representation of different mathematical operators implemented by the proposed metasurface, thereby allowing the calculus operations to be performed in the far-field scattering pattern. For a metasurface with continuous electromagnetic response, the near-field response $r(x)$ and the corresponding one-dimensional far-field scattering pattern $R(\theta)$ can be represented as:

$$R(\theta) = \int_{x_{min}}^{x_{max}} r(x) e^{-j\frac{2\pi x \sin\theta}{\lambda}} dx \tag{2}$$

where $\lambda$ is the wavelength of the incident EM waves. For the proposed metasurface adopted in this work, the EM response is discretized in space. Accordingly, the discrete near-field response $r(x_i)$ and the scattering pattern $R(\theta)$ can be further represented as:

$$R(\theta) = \sum_{i=1}^{N} r(x_i) e^{-j\frac{2\pi x_i \sin\theta}{\lambda}} \tag{3}$$

where $x_i$ is the center of the meta-atom and $N$ is the number of meta-atom.

When conducting the mathematical calculations, the proposed STCM is illuminated by a monochromatic EM wave that generates a near-field distribution $f(x)$ according to the far-field distribution $F(\theta)$. The metasurface generates a phase and amplitude distribution $r(x_i)$ at a specific harmonic to represent a mathematical operator. The output of the mathematical calculation $F_1(\theta)$ can be obtained by measuring the resulting far-field pattern, where $F_1(\theta)$ can be expressed as:

$$F_1(\theta) = \sum_{i=1}^{N} f(x_i) r(x_i) e^{-j\frac{2\pi x_i \sin\theta}{\lambda}} \tag{4}$$

Owing to the Fourier transform relation, the multiplication of $f(x)$ and $r(x_i)$ in the near-field corresponds to the convolution of $F(\theta)$ and $R(\theta)$ in the far field. Therefore, if the phase and amplitude distribution $r(x_i)$ is optimized by a calculus operator, the metasurface can perform the calculus operations for the incident EM waves in the far-field scattering pattern.

## Theoretical results and numerical simulations

To implement practical STCM-based calculus operations, we adopt a 2-bit programmable metasurface as the hardware platform. The configuration of the employed metasurface and its EM reflection response are shown in **Figure 2**, while the detailed structure of the meta-atom can be found in our previous work [39]. The reflection response of the meta-atom is controlled by two PIN diodes that are embedded in the meta-atom. For convenience, the four different combinations of PIN-diode states are defined as four distinct binary coding states from 00 to 11. The coding states and the measured reflection responses of the meta-atom at 10.3 GHz are summarized in Table 1.

Table 1. The coding states and the reflection responses of meta-atom

| Coding states | PIN diode-1 | PIN diode-2 | Amplitude (dB) | Phase (°) |
|---|---|---|---|---|
| 00 | OFF | OFF | -0.06 | 12° |
| 01 | ON | OFF | -2.86 | 102° |
| 10 | OFF | ON | -3.25 | 186° |
| 11 | ON | ON | -0.98 | 289° |

*The reflection responses are measured at 10.3 GHz

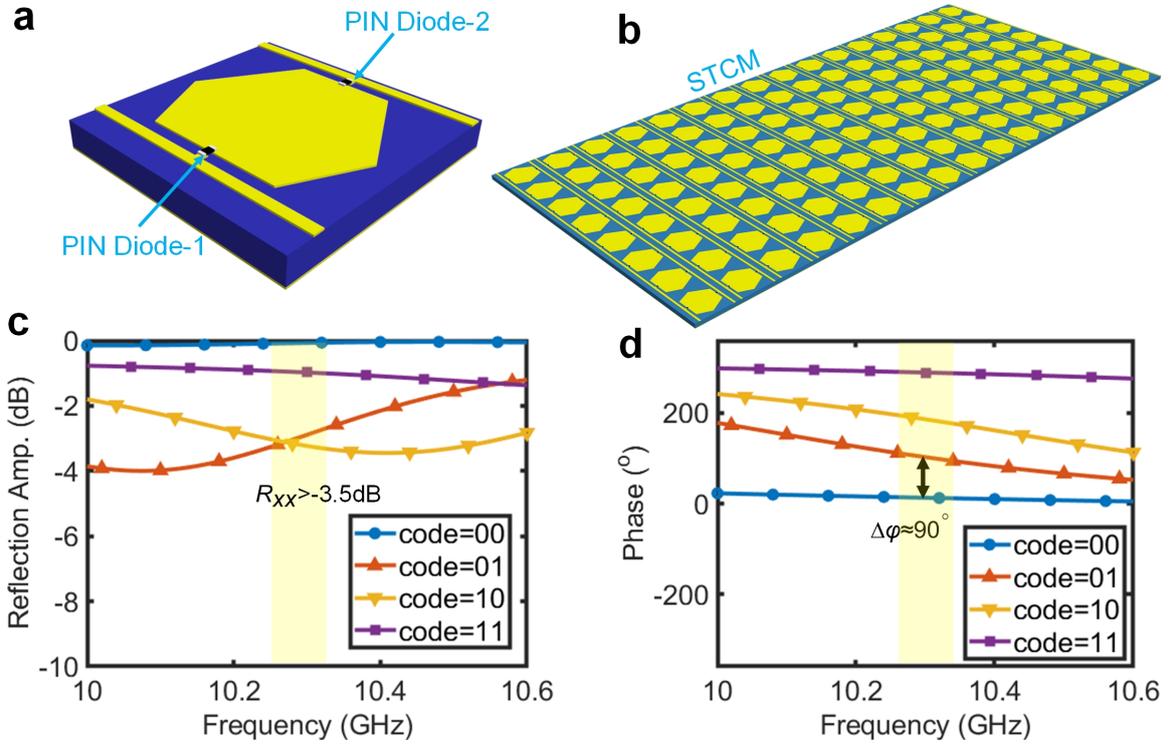

**Figure 2. The structure and reflection responses of the STCM meta-atom.** a) The structure of the meta-atom. b) The schematic view of the realized 2-bit STCM prototype with 16 columns. c) The reflection amplitudes of the meta-atom with different coding states. d) The reflection phases of the meta-atom with different coding states.

To further evaluate the performance of meta-atom, the measured results of the adopted meta-atom in the frequency band 10~10.6 GHz are shown in Figures 2c and 2d. These results demonstrate that the proposed meta-atom has four different phase responses with an approximately 90° phase increment, while a stable reflection amplitude higher than -3.5 dB in four different coding statuses from 10~10.6 GHz. The 2-bit coding performance confirms that the proposed metasurface can serve as an ideal hardware platform for implementing calculus operations.

To verify the concept of the STCM-based calculus platform, we conduct simulations on a prototype with 16 columns that are controlled independently, as shown in Figure 2b. Based on the Fourier transform relation between the time-coding sequences and harmonic responses, the time-coding sequences can be obtained by inverse Fourier transform of a given harmonic response sequence if the amplitude and phase responses of the meta-atom can be manipulated arbitrarily. However, for the given 2-bit programmable metasurface, the meta-atom only has four limited reflection responses. Hence, for a given periodic time-coding sequence of length $L$, the optimization of time-coding sequence can be regarded as a search problem in a finite state space. To accelerate the solution of optimal time-coding sequence, Genetic Algorithm (GA) is employed for the optimization process.

To show the effectiveness of GA in optimizing the coding sequences, we conduct the differential and integral operations at the first-order harmonic, respectively. When conducting the differential operation in the first-order harmonic, the constraint condition $err_k$ for the space-time coding sequence of different coding elements in STCM with $L$ independent controllable coding elements is expressed as

$$err_k = \left| \left(k - \frac{N+1}{2}\right) jA - \sum_{n=1}^{L} c_{kn} e^{-j\frac{2n\pi}{L}} \right| \qquad (5)$$

where $k$ is the number of the coding element, and $A$ is a constant value for controlling the harmonic amplitude. When $A$ is specified as 0.0875 and $err_k$ is minimized, the optimization results of STCM with 16 coding columns are presented in **Figure 3a**, and the near-field harmonic responses of each meta-atom are given in Figures 3c and 3d. To illustrate the performance of the differential operation, we simulate the scattering pattern of the proposed metasurface illuminated by normally incident EM waves. Figure 3d shows the scattering pattern of a metal plate with the same area as the STCM calculus platform, while Figure 3e presents the simulation results of the STCM calculus platform performing the differential operation, alongside the theoretical differential results for the incident wave reflected from a metal plate of identical size.

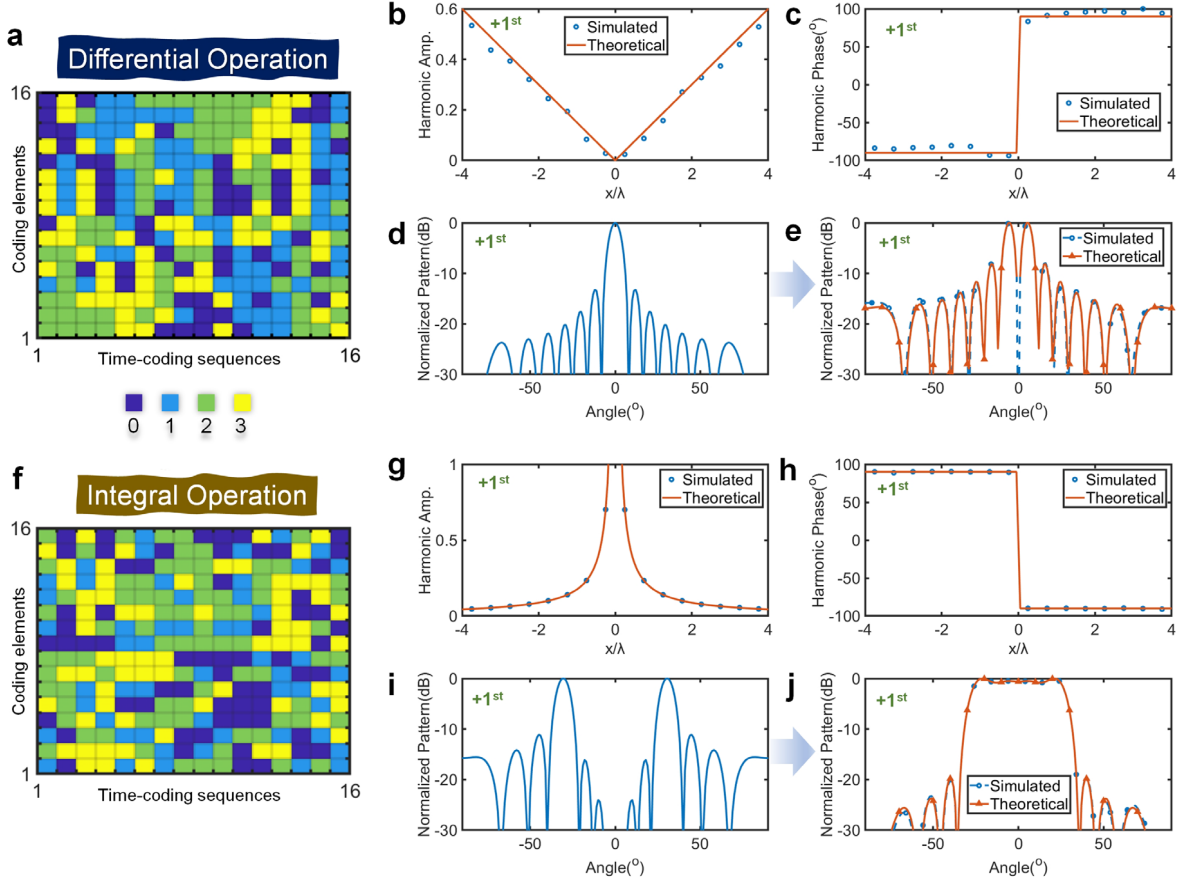

**Figure 3. Simulated results of performing differential or integral operations at the +1st harmonic.** a) Space-time coding sequences of STCM for differential operation at the +1st harmonic. b) The amplitude responses of different coding elements in the differential operation. c) The phase responses of different coding elements in the differential operation. d) The normalized scattering pattern of the metal plate with the same area illuminated by a normally incident plane wave. e) The normalized scattering pattern of the metasurface illuminated by the normally incident plane wave when conducting the differential operation. f) Space-time coding sequences of STCM for integral operation. g) The amplitude responses of different coding elements in the integral operation. h) The phase responses of different coding elements in the integral operation. i) The normalized scattering pattern of the metal plate with the same area illuminated by differential incident waves at ±30°. j) The normalized scattering pattern of the metasurface illuminated by differential incident waves at ±30° when conducting the integral operation.

For the integral operation, the constraint condition $err_k$ for the space-time coding sequence is defined as

$$err_k = \left| \frac{jA}{k-\frac{N+1}{2}} - \sum_{n=1}^{L} c_{kn} e^{-j\frac{2n\pi}{L}} \right| \qquad (6)$$

When $A$ is specified as 0.35 and $err_k$ is minimized, the optimization results of the metasurface with 16 coding columns are shown in Figure 3f, and the near-field harmonic responses of meta-atom are presented in Figures 3g and 3h. To illustrate the performance of the integral operation, we simulate the scattering pattern of the proposed metasurface illuminated by differential incident waves at the angle of ±30°, where the incident waves from 30° and -30° have a phase difference of 180°. Figure 3i shows the scattering pattern of the metal plate with the same area as the metasurface, and Figure 3j shows the simulated results of the STCM calculus platform performing the integral operation and the theoretical integral results of the incident waves that are reflected by the metal plate of identical size.

In the harmonic responses, the good agreement between the simulated results and the theoretical values demonstrates the effectiveness of GA in optimizing the space-time coding sequences, and the unparalleled degree of freedom for STCM in harmonic manipulations. In the far-field scattering patterns, the STCM calculus platform can successfully split a single beam through differential operation and convert differential incident waves into a fan-shaped beam through the integral operation. The good consistency between theoretical differential and integral beams and the far-field scattering pattern of STCM demonstrates the feasibility of directly performing calculus operations on spatial EM waves using metasurfaces. These results highlight the broad application prospects for metasurfaces in EM manipulation and signal processing.

To fully use the EM manipulation capability of STCM and broaden its potentials in EM calculus operations, a space-time coding sequence is optimized to simultaneously perform the differential operation at the first harmonic and the integral operation at the second harmonic. The corresponding constraint condition $err_k$ for this space-time coding sequence is defined as

$$err_k = \sqrt{\left|\left(k - \frac{N+1}{2}\right)jA_1 - \sum_{n=1}^{L} c_{kn} e^{-j\frac{2n\pi}{L}}\right|^2 + \left|\frac{jA_2}{k - \frac{N+1}{2}} - \sum_{n=1}^{L} c_{kn} e^{-j\frac{4n\pi}{L}}\right|^2} \qquad (7)$$

To ensure the feasibility of the coding optimization, the following constraints are imposed on $A_1$ and $A_2$:

$$\sqrt{\left[\left(k - \frac{N+1}{2}\right)A_1\right]^2 + \left(\frac{jA_2}{k - \frac{N+1}{2}}\right)^2} < 1 \qquad (8)$$

When $A_1$=0.075 and $A_2$=0.35, the simulated results of STCM that simultaneously conducts the differential and integral operations in the first and second harmonics are presented in **Figure 4**. Compared with the results that separately implement the differential and integral operations in the first harmonic (see Figure 3), the amplitude and phase responses of the first and second harmonics for simultaneously differential and integral operations in Figures 4c-4f exhibit larger deviations from the theoretical values. However, the simulated results still verify the feasibility to conduct different calculus operations in different harmonics without increasing the encoding length. Moreover, the simulation points with relatively large errors typically correspond to small amplitudes, reducing their impact on the calculus operations.

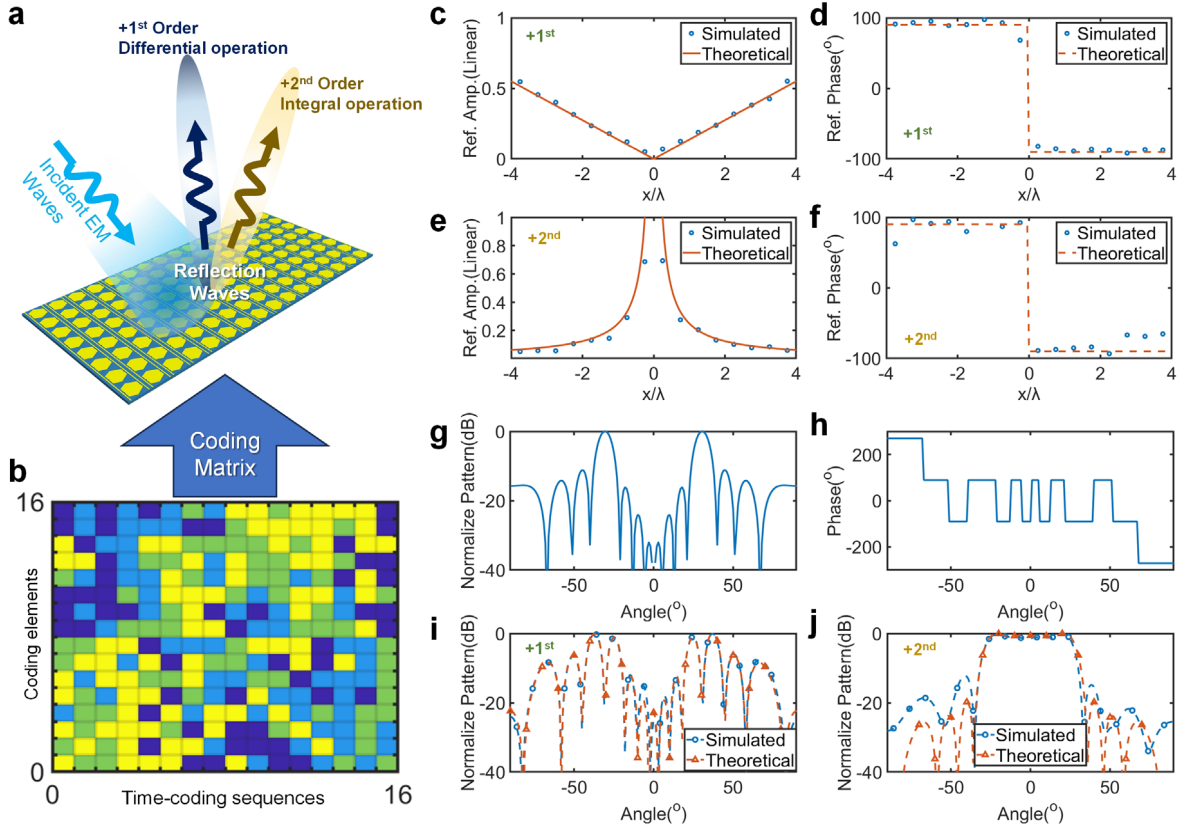

**Figure 4. Simulated results of simultaneously performing differential and integral operations at the +1st and +2nd harmonics, respectively.** a) Schematic diagram of STCM that simultaneously performs the differential and integral operations. b) Space-time coding sequences for performing the calculus operations at two harmonics. c) The amplitude responses of different coding elements at the +1st harmonic. d) The phase responses of different coding elements at the +1st harmonic. e) The amplitude responses of different coding elements at the +2nd harmonic. f) The phase responses of different coding elements at the +2nd harmonic. g) The normalized scattering pattern of the metal plate with the same area illuminated by differential incident waves at ±30°. h) The phase distribution of the metal plate with the same illumination. i) The normalized scattering pattern of STCM at the +1st harmonic with the same illumination when conducting the differential operation. j) The normalized scattering pattern of STCM at the +2nd harmonic with the same illumination when conducting the integral operation.

To further evaluate the performance of STCM calculus platform, numerical simulations are conducted on scattering patterns of the first and second harmonics when the metasurface with the predesigned space-time coding sequence is illuminated by the differential incident waves at ±30°. For comparison, the reflection amplitude and phase of a metal plate with the same area under the same illumination are shown in Figures 4g and 4h, while the far-field scattering patterns of the first and second harmonics are presented in Figures 4i and 4j. The simulation results are in good consistency with the theoretical value, strongly supporting the application prospects of STCM for multi-functional calculus operations.

To further verify the feasibility of STCM to directly perform the calculus operations on EM waves, we consider a practical application for image processing, where performing the differentiation on an image can enhance its edges, thereby facilitating the edge detection. To demonstrate the edge enhancement using the STCM calculus platform, numerical simulations are conducted with STCM consisting of 16×16 independently controllable meta-atoms, in which a space-time coding sequence with a period of 16 time slots is employed. During the coding optimization, the constraint condition $err_{kl}$ for the space-time coding sequence is defined as

$$err_{kl} = \left| \left( \sqrt{k^2 + l^2} - \frac{N+1}{2} \right) e^{-j \tan^{-1}\left( \frac{k - \frac{N+1}{2}}{l - \frac{N+1}{2}} \right)} A - \sum_{n=1}^{L} c_{kln} e^{-j \frac{2n\pi}{L}} \right| \quad (9)$$

where $k$ and $l$ are the row and column numbers corresponding to the meta-atom.

The simulated results of the amplitude and phase responses for the optimized space-time coding sequence with $A_{max}$=0.052 at the first harmonic are shown in **Figures 5a** and **5b**. These results indicate that the reflection amplitude forms a conical distribution and the phase forms a vortex distribution at the first harmonic, corresponding to the Fourier transform of a two-dimensional (2D) first-order differential operator. To verify the edge enhancement of the STCM-based calculus platform, simulations are conducted with three different structured EM waves, which form three different images, including 'S', 'E', and 'U', in the far-field scattering patterns of a PEC plate with the same area as the metasurface, as shown in Figures 5c, 5e, and 5g. The corresponding simulation results of the structured EM incident waves reflected by the STCM performing 2D differential operation are shown in Figures 5d, 5f, and 5h. In the simulation, the pitch and azimuth angles are normalized to the range of -1 to 1 through a sine function. The results demonstrate that a metasurface with 16×16 meta-atoms can successfully implement a 2D differential operation towards incident waves and realize edge enhancement, indicating the significant potential of the STCM-based calculus platform for image-processing applications.

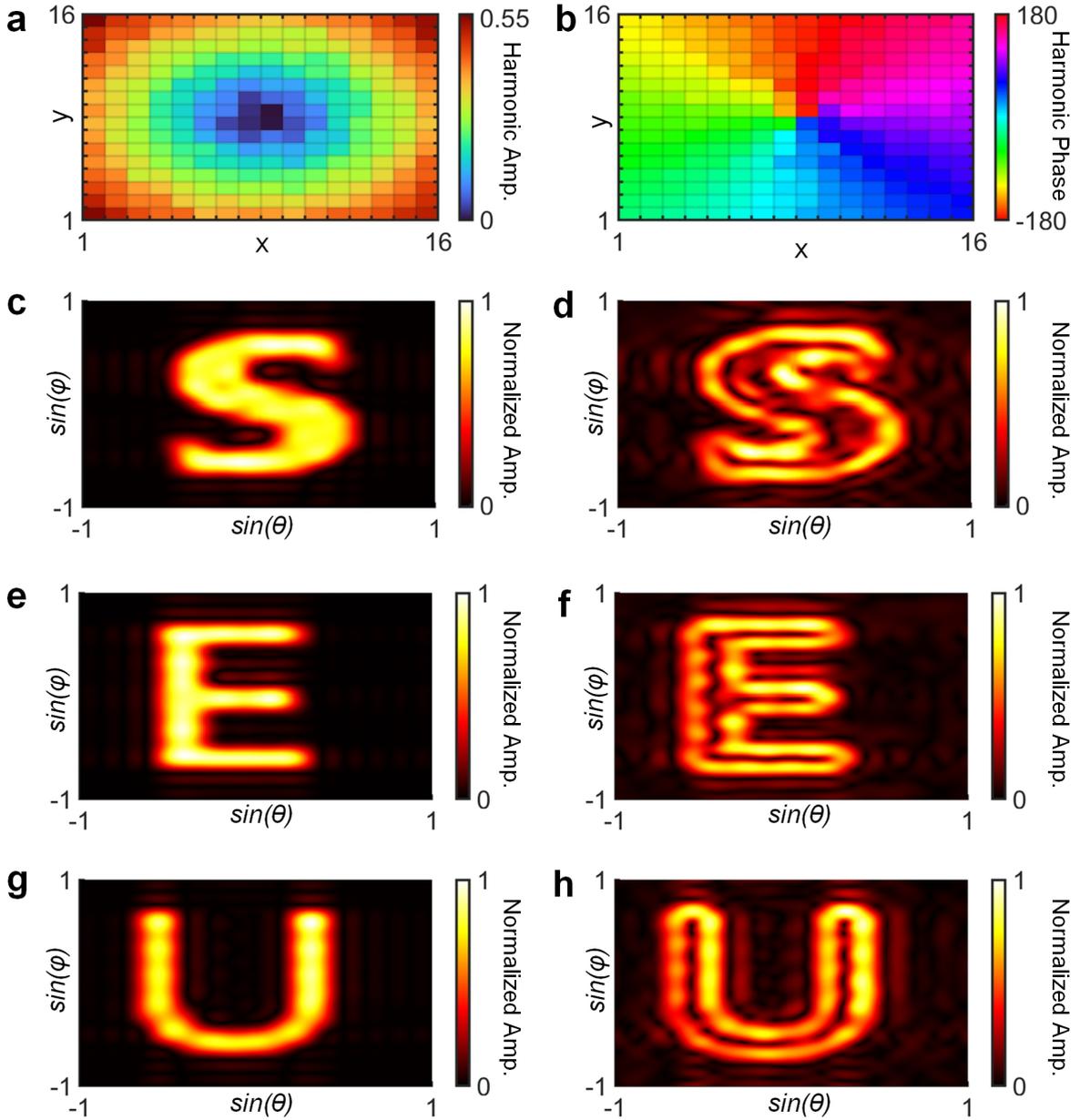

**Figure 5. Simulated results of two-dimensional differential operations for edge enhancement at the +1st harmonic.** a) The amplitude distribution of STCM with 16×16 meta-atoms. b) The phase distribution of STCM with 16×16 meta-atoms. c-h) The scattering patterns of the metal plate (c, e, g) and STCM calculus platform (d, f, h) when conducting the edge enhancement of 'S', 'E', and 'U', respectively.

## Experimental validations

The simulated results verify the feasibility of performing the calculus operations using STCM. To realize a practical STCM prototype for constructing a calculus platform and evaluate its performance, we fabricated a 16×8 reflection-type programmable metasurface using a printed circuit board (PCB). The prototype was measured in a far-field microwave anechoic chamber to verify the function and performance, with the measurement setup shown in **Figure 6a**. To enable the harmonic manipulations, we apply the control signals with a modulation frequency of 1 MHz dynamically to the metasurface via an FPGA control module. A signal generator connected to a transmitting antenna is used to generate a monochromatic incident EM wave at 10.3 GHz. The far-field scattering pattern is received by a receiving antenna while the rotary platform rotates from -90° to 90°.

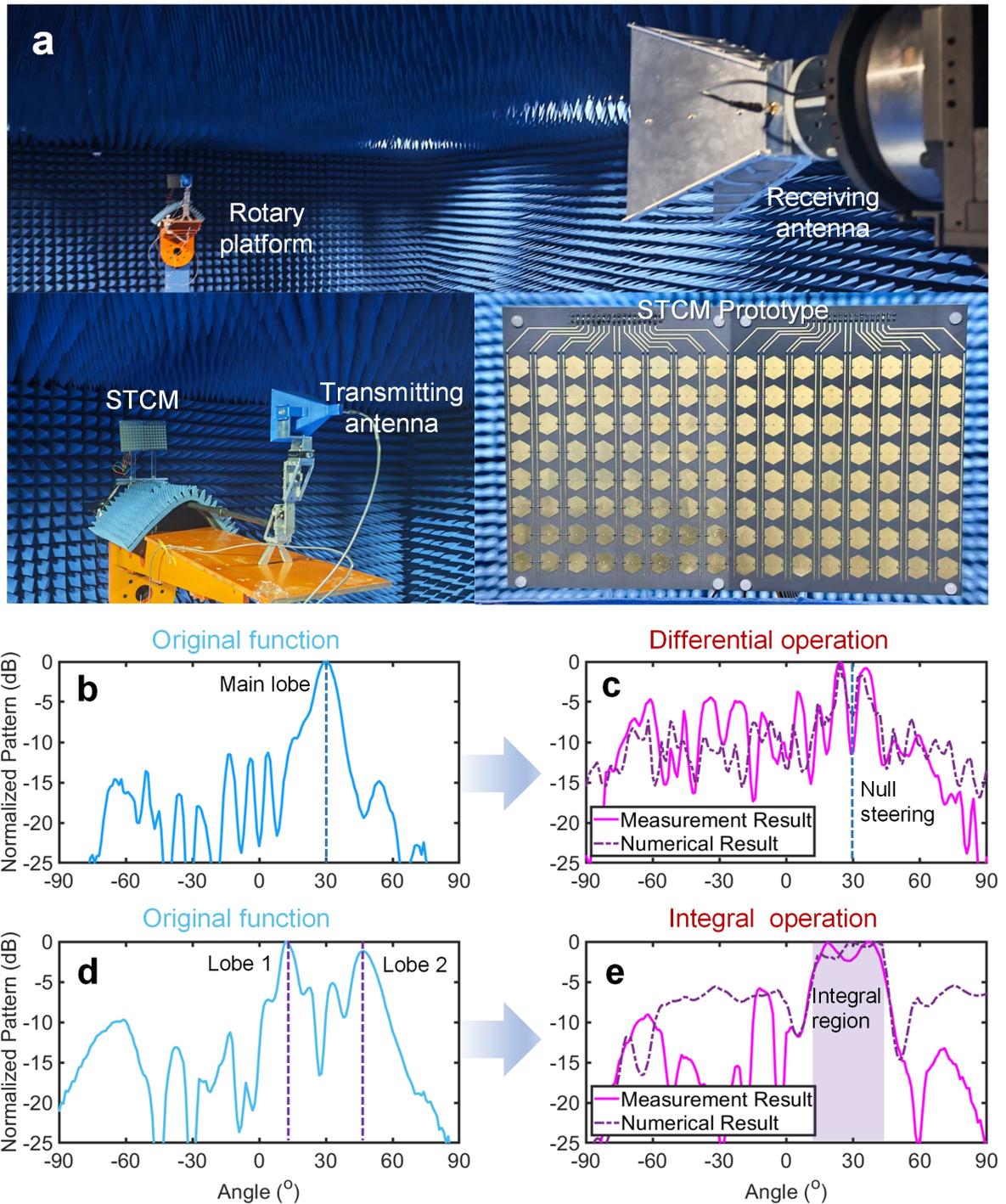

**Figure 6. Experiment setup and measurement results of the STCM-based calculus operations.** a) The experimental setup and measurement environment of the STCM-based calculus platform. b) The scattering

pattern of the original function for the differential operation. c) The measured and numerical calculation results of the differential operations with the original function 1. d) The scattering pattern of the original function for the integral operation. e) The measured and numerical calculation results of the integral operations with the original function 2.

To avoid the blockage of the transmitting antenna, a phase gradient is incorporated into the optimization of the space-time coding sequence, resulting in the reflection beam deflected by 30°. For verification of the differential operation, the constraint condition $err_{k0}$ for the space-time coding sequence for column $k$ in the original function is defined as

$$err_k = \left| e^{-j2\pi\frac{kd}{\lambda}\sin\frac{\pi}{6}} - \sum_{n=1}^{L} c_{kn} e^{-j\frac{2n\pi}{L}} \right| \tag{10}$$

where $d$ is the width of each column, and $\lambda$ is the operation wavelength. For the differential operation, the constraint condition $err'_{k1}$ for the space-time coding sequence for column $k$ is defined as

$$err'_k = \left| \left(k - \frac{N+1}{2}\right) j A_{max} e^{-j2\pi\frac{kd}{\lambda}\sin\frac{\pi}{6}} - \sum_{n=1}^{L} c_{kn} e^{-j\frac{2n\pi}{L}} \right| \tag{11}$$

Theoretically, the far-field scattering pattern of the original function at the first harmonic forms a single beam pointing at 30°, whereas the scattering pattern of STCM performing the differential operation exhibits a split beam with a null at 30° and two peaks located at the rising and falling edges of the original function. To evaluate the performance of the differential operation, numerical differential result of the original function 1 and measured result of the scattering pattern of STCM implementing the differential operation are shown in Figure 6c. Although discrepancies are observed in the sidelobe region, both numerical and measured results present excellent differential performance in the main-lobe region, showing fairly agreement with the theoretical predictions.

For the integral operation, since generating a differential incident wave is challenging, the original function is realized by adjusting the harmonic response of the space-time coding sequence. To form a pair of beams with a phase difference of 180°, a 1-bit phase coding sequence $C_0$ is employed to constrain the harmonic response of the metasurface element, where $C_0$ = 0000111100001111. Similarly, to avoid the blockage of the transmitting antenna, a phase gradient capable of deflecting reflection beams by 30° is incorporated into the optimization of the space-time coding sequence in the original function and integral operation. The constraint condition $err_k$ for optimizing the original function is defined as

$$err_k = \left| e^{-j(2\pi\frac{kd}{\lambda}\sin\frac{\pi}{6} - C_0(k)\pi)} - \sum_{n=1}^{L} c_{kn} e^{-j\frac{2n\pi}{L}} \right| \tag{12}$$

while the constraint condition $err'_{kl}$ for optimizing the integral operation is defined as

$$err'_k = \left| \frac{jA_{max}}{k - \frac{N+1}{2}} e^{-j(2\pi\frac{kd}{\lambda}\sin\frac{\pi}{6} - C_0(k)\pi)} - \sum_{n=1}^{L} c_{kn} e^{-j\frac{2n\pi}{L}} \right| \tag{13}$$

Theoretically, the scattering pattern of the original function at the first harmonic forms two symmetrical beams around 30°, while the scattering pattern of the metasurface performing the integral operation produces a beam with stable amplitude around 30°. The measurement results are presented in Figures 6d and 6e. Specifically, the measured scattering pattern of the original function in the integral function case exhibits two lobes with a 180° phase difference at 13° and 47°. In contrast, the scattering pattern resulting from the integral operation maintains a stable scattering pattern with a normalized amplitude higher than -3dB over the angular range from 13° to 43°. The numerical integral results of the original function are also presented in Figure 6e. Although discrepancies between the measured and numerical results are observed in the sidelobe region, the numerical integration exhibits satisfactory consistency with the measured results in the integral region, while the power levels in sidelobes are much lower than those in the integral region. Overall, the measured results around 30° show good consistency with the theoretical analysis, confirming the feasibility of implementing the integral operation using STCM.

As a summary, despite the measurement error discrepancies arising from the unideal conduction rate of PIN-diodes and the inevitable differences between the simulated and measured performance of the metasurface, the experimental results obtained from the STCM-based calculus platform demonstrate its powerful ability to perform the calculus operations in the EM space.

## Conclusion

We propose a programmable STCM-based calculus operation platform. A carefully designed reflection-type metasurface with a 2-bit phase response and stable reflection amplitude provides a robust hardware foundation for the EM wave-based calculus platform. By using different time-coding sequences, the reflection phase and amplitude at specific harmonics can be flexibly and precisely controlled, enabling STCM to emulate the differential and integral operators in the Fourier-transform domain. Leveraging the Fourier-transform relationship between the near-field response and far-field scattering pattern, the proposed STCM realizes the direct calculus operations in the EM space. Compared with previous approaches, the STCM calculus platform exhibits versatile, reprogrammable, and multifunctional operations with a low profile. Moreover, the programmable nature of STCM offers significant potentials for performance enhancement and functional extension. Good agreement among theoretical analysis, numerical simulations, and experimental measurements confirms the feasibility of STCM to perform calculus operations in EM space, highlighting the broad application prospects of STCM in wireless communication, remote sensing, and signal processing.

## Acknowledgements

This work was supported by the National Natural Science Foundation of China (62288101, 62101123, U22A2001, and 62201136), the National Key Research and Development Program of China (2023YFB3811504), the Jiangsu Province Frontier Leading Technology Basic Research Project (BK20210210, BK20212002), the Xiaomi Foundation, the 111 Project (111-2-05), the SEU Innovation Capability Enhancement Plan for Doctoral Student (CXJH_SEU 26088), and the Fundamental Research Funds for the Central Universities (2242023K5002).